\documentclass[conference,a4paper]{IEEEtran}
\usepackage{multirow}
\usepackage{amssymb}
\usepackage{amsmath}
\usepackage{epsfig}
\usepackage{url}
\usepackage{cite}
\usepackage{verbatim}
\usepackage{algorithm, algorithmic}
\usepackage{color}

\usepackage{lipsum}
\usepackage{float}
\restylefloat{figure}

\def\bnu{\mbox{\boldmath $\nu$}}

\def\kmer{\mbox{$k$mer~}}
\def\kmers{\mbox{$k$mers~}}

\newcommand{\cG}{\mathcal{G}}

\newcommand{\Msg}{\boldsymbol{s}}

\newcommand{\setRead}{\mathcal{R}}

\newcommand{\baseCalls}{\boldsymbol{x}}
\newcommand{\qScores}{\boldsymbol{y}}

\newcommand{\statespace}{\mathcal{K}}

\newcommand{\nbhd}[1]{\mathcal{N}^d(#1)}
\newcommand{\paramSpace}{\boldsymbol{\theta}}
\newcommand{\substr}[3]{#1{\scriptstyle [#2\dots#3]}}
\newcommand{\indexof}[2]{#1{\scriptstyle [#2]}}

\newcommand{\transDist}[2]{p(#2|#1)}

\newcommand{\vs}{\vspace{-0.0in}}
\renewcommand{\vss}{\vspace{-0.05in}}
\begin{document}

\sloppy

\title{PREMIER - PRobabilistic Error-correction using Markov Inference in Errored Reads}

\author{
    \IEEEauthorblockN{Xin Yin\IEEEauthorrefmark{1}, Zhao Song\IEEEauthorrefmark{2}, Karin Dorman\IEEEauthorrefmark{1}\IEEEauthorrefmark{3} and Aditya Ramamoorthy\IEEEauthorrefmark{2}}
    \IEEEauthorblockA{\IEEEauthorrefmark{1}Dept. of Statistics,
    Iowa State University,
    Ames, IA 50011\\}
    \IEEEauthorblockA{\IEEEauthorrefmark{2}Dept. of Electrical \& Computer Eng.
    Iowa State University,
    Ames, IA 50011\\}
    \IEEEauthorblockA{\IEEEauthorrefmark{3}Dept. of Genetics, Development \& Cell Biology
    Iowa State University,
    Ames, IA 50011\\
    \{xinyin, zhaosong, kdorman, adityar\}@iastate.edu}
}
\maketitle

\begin{abstract} THIS PAPER IS ELIGIBLE FOR THE STUDENT PAPER AWARD.
In this work we present a flexible, probabilistic and reference-free method of error correction for high throughput DNA sequencing data. The key is to exploit the high coverage of sequencing data and model short sequence outputs as independent realizations of a Hidden Markov Model (HMM). We pose the problem of error correction of reads as one of maximum likelihood sequence detection over this HMM. While time and memory considerations rule out an implementation of the optimal Baum-Welch algorithm (for parameter estimation) and the optimal Viterbi algorithm (for error correction), we propose low-complexity approximate versions of both. Specifically, we propose an approximate Viterbi and a sequential decoding based algorithm for the error correction. Our results show that when compared with Reptile, a state-of-the-art error correction method, our methods consistently achieve superior performances on both simulated and real data sets.
\end{abstract}

\section{Introduction}
\vs
DNA sequencing is the process of finding the identity and order of nucleotides or bases, adenine ($A$), guanine ($G$), cytosine ($C$) and thymine ($T$), in DNA molecules.
It is used widely in biological and medical research to determine the genomes of diverse organisms ranging from microbes to humans.
In recent years the advent of low cost, high throughput DNA sequencing~\cite{Metzker2010} has made it feasible to sequence multiple individuals, even entire populations of organisms.
This technological advance may be the key to achieving truly personalized medicine, and several other grand goals in biology.

The summed length of the DNA molecules that need to be sequenced can vary from a few thousand bases to hundreds of gigabases. 
Sequencing operates by breaking the DNA double helix molecules at random locations and generating incomplete ``reads'' that start at one end of the resulting fragments and read contiguous bases along one of the two strands.
Read lengths are quickly increasing, but vary from thirty base pairs (bp) in the past to over a thousand bp on some platforms~\cite{Quail2012}.
%

A critical issue in DNA sequencing is the elevated error rate in reads from the current technology~\cite{Chan2009}.
While the throughput rate is high, substitution errors, \textit{e.g.},~base $A$ called as $C$, and insertion/deletion errors where spurious bases are included or valid bases are left out, are frequent.
Errors in reads pose a serious problem for downstream uses of sequence data, including sequence assembly~\cite{Salzberg2012}, where the full-length sequence is inferred from the short reads, and variant identification~\cite{Wilm2012}, for detecting genetic heterogeneity in a population.

The primary approach for dealing with errors is to capitalize on the high throughput of the sequencing technology.
Such an excess of fragments are sequenced that each base in the DNA molecules is covered by multiple reads.
However, because the starting location of each read is random, there is no alignment information to indicate which reads cover a particular base.
This lack of alignment information makes the problem different from classical error correction \cite{lincostello}.
Indeed if alignment information were available, the problem would roughly reduce to the decoding of a repetition code. 


Error correction of noisy reads has received significant attention in the bioinformatics community in recent years~\cite{Yang2012}.
We briefly review the methods most closely related to our proposed method.
Many methods begin by counting the occurrence of all $k$mers in the reads.
A \kmer is a substring of length $k$.
Euler~\cite{Chaisson2009} corrects a read via the smallest set of corrections that make all \kmers in the read \textit{common}, \textit{i.e.}~have high occurrences.
Hammer~\cite{Medvedev2011} identifies cliques by linking similar $k$mers, then corrects all members to the clique's consensus $k$mer.
FreClu~\cite{Qu2009} corrects full-length reads if it finds a significantly more frequent read differing at just one position. 
Quake~\cite{Kelley2010} iteratively corrects bases by maximizing a posterior probability of the true sequence given the observed read until all \kmers are common.
Two other methods use a probability model to correct a read~\cite{Wijaya2009} or \kmer \cite{Yang2011} to the most likely true sequence.
In all these methods the focus must turn to \kmers when the read length is long to guarantee sufficient repetition to distinguish error and true bases.
So as read lengths increase, even read-based methods must become $k$mer-based.
All $k$mer-based methods ignore the fact that \kmers are dependently read as contiguous substrings within reads.
Moreover, while some allow arbitrarily complex error models, either all error parameters must be provided \textit{a priori} or the parameter estimation procedure is \textit{ad hoc}. 

The work of \cite{motahariBT2012}, modeled a genome as the output of a discrete memoryless source and determined the coverage levels required to guarantee correct sequence assembly under a noiseless read process. Approaches based on statistical modeling of the sequencing process have been used \cite{DasV12, WuV12} for basecalling, but not for error correction of reads.

\noindent \underline{{\it Main contributions.}} In this work we address the problem of correcting errors in noisy reads from a signal processing and error control coding viewpoint. We consider reads from the Illumina DNA sequencer that is known to exhibit substitution errors (but essentially no insertion/deletion errors)~\cite{Ledergerber2011a}.
We demonstrate that Illumina reads can be modeled as symbols emitted from 
a Hidden Markov Model (HMM).
To overcome the unmanageably large state space, we use constraints and penalties to estimate the HMM parameters.
Given the parameters of this HMM, we pose the problem of error correction of reads as a maximum likelihood sequence detection problem.
While time and memory considerations rule out an implementation of the optimal Baum-Welch algorithm (for parameter estimation) and the optimal Viterbi algorithm (for error correction), we propose low-complexity approximate versions of both. This approach is successful in identifying many errors. 
In addition we propose a sequential decoding \cite{Fano} algorithm that achieves even better performance.
Our results on {\it real}, publicly available sequencing data for the \textit{E. coli} genome demonstrate a $9$\% improvement in error correction rates over a current state of the art technique.
\vss
\section{Problem Formulation}
\vs
Let $\cG$ denote the genome to sequence.
It is a $4$-ary sequence of length $|\cG|$, where each letter is in $\Omega = \{A,C,G,T\}$; we call these letters bases or nucleotides.
Sequencing operates by breaking the genome into fragments, from which length $L$ reads are made.
The sequencer has access to multiple copies of $\cG$ and produces up to billions of reads. 
The starting point of the fragment within $\cG$ is random and unknown.
The sequencer processes the fragment and outputs a read $(\baseCalls, \qScores)$, where $\baseCalls$ is the sequencer's best guess of $L$ bases in the fragment and $\qScores$ are the corresponding quality scores; the quality scores are discrete measures of confidence in the base calls $\baseCalls$.
A given base location will typically be covered by multiple reads. Let $N$ denote the total number of reads obtained in this process. The coverage level is defined to be $NL/|\cG|$.
\vss
\subsection{HMM modeling}
\label{subsec:HMM}
\noindent
We model the sequencer as a Hidden Markov Model (HMM); each read is an independent realization of the HMM.
Given $\Msg$, the unknown true read of length $L$, we define $\indexof{\Msg}{i}$ as the $i$-th character of $\Msg$ and $\substr{\Msg}{i}{j}$ as the substring from position $i$ to $j$ (both $\indexof{\Msg}{i}$ and $\indexof{\Msg}{j}$ are included).
Let $\Msg_t = \substr{\Msg}{t-k+1}{t}$ be the $t$-th \emph{state}. In the discussion below we will also refer to the states as $k$mers.
Similarly, $\baseCalls_t = \substr{\baseCalls}{t-k+1}{t}$ will be referred to as the $t$th observed \kmer and $\qScores_t =  \substr{\qScores}{t-k+1}{t}$ will denote the corresponding quality scores. We model the sequencer as transitioning between the states $\Msg_{t-1}$ to $\Msg_{t}$.
On the $t$th transition it emits the output $(\indexof{\baseCalls}{t},\indexof{\qScores}{t})$.

To specify the model completely, we need to define:
\begin{itemize}
\item State space $\statespace$, where $|\statespace| \leq 4^k$.
\item Transition distribution $\transDist{\Msg_t}{\indexof{\Msg}{t+1}}$, where
    \[
        \sum_{\beta\in \Omega} \transDist{\boldsymbol{\alpha}}{\beta} =  1, \quad \forall \boldsymbol{\alpha} \in \statespace.
    \]
\item Emission distribution $f_t(\indexof{\baseCalls}{t}, \indexof{\qScores}{t}\mid\Msg_t)$ with
    \begin{align}
        & f_t(\indexof{\baseCalls}{t}, \indexof{\qScores}{t}\mid\Msg_t) = q_t(\indexof{\qScores}{t} \mid \indexof{\baseCalls}{t}, \Msg_t) g_t(\indexof{\baseCalls}{t}\mid\Msg_t),
    \end{align}
where we assume the following simple forms.
\begin{align}
    & q_t(\indexof{\qScores}{t}\mid\indexof{\baseCalls}{t}, \Msg_t)  =
        \begin{cases}
            q_{t0}(\indexof{\qScores}{t}) & \indexof{\baseCalls}{t} = \indexof{\Msg}{t}, \\
            q_{t1}(\indexof{\qScores}{t}) & \indexof{\baseCalls}{t} \neq \indexof{\Msg}{t}, \text{~and}
        \end{cases} \nonumber\\
    \label{eqn:emitDist}
    & g_t(\indexof{\baseCalls}{t}\mid\Msg_t) = 1\{\Msg_t \in \nbhd{\baseCalls_t}\}g_t(\indexof{\baseCalls}{t}\mid\indexof{\Msg}{t}), \\
    & \hspace{0.1in}\mbox{with } \sum_{\beta\in\Omega} g_t(\beta \mid \beta') = 1, \quad \forall \beta'\in\Omega. \nonumber
\end{align}
\end{itemize}
Our modeling philosophy is guided by the following considerations.
It is well recognized that nucleotides in genomes display strong local dependence, and Markov models, like the one we use for $\Msg$, have long been used to model this dependence~\cite{Picardi2010}.
Like most error correction correction methods, we start with a simple error model.
Both $g_t(\beta\mid \beta')$, the probability of (mis)reading base $\beta'$ as $\beta$, and $q_{tj}(q), j=0,1$, the quality score probability mass functions, depend on position $t$.
We expect that $q_{t0}(q)$ is shifted right of $q_{t1}(q)$, for all $t$, because the sequencer should assign higher quality scores to error free bases.
We defer discussion of the role of the indicator function $1\{\cdot\}$  in eq. (\ref{eqn:emitDist}) (see text near eq.~(\ref{def:kmer_nbhd}) below).
%

The choice of \kmer length $k$ and state space $\statespace$ are guided by several considerations.
In principle, given $k$, one could choose all possible $4^k$ states as $\statespace$, but even for moderate $k$ (around 15), such $\statespace$ is too big.
Thus, for a given set of reads, we restrict $\statespace$ to contain only observed $k$mers. 
Even though $\statespace$ includes erroneous $k$mers, 
we hope to identify them during estimation of the HMM.
The choice of $k$ depends on two conflicting requirements.
On the one hand, we want an accurate model.
If $k$ is too small, say $k = 3$, then each $k$mer, say $ATC$, will exist in several locations in the original $\cG$. 
Our model will tend to ``correct'' uncommon downstream bases to common downstream bases.
For example, if $ATCG$ occurs twice and $ATCT$ occurs once, then true read $ATCT$ may be erroneously corrected to $ATCG$. 
On the other hand, very large $k$ (though it cannot exceed the read length $L$), may lead to decreased \kmer coverage and eventually an overparameterized, inestimable model.
Thus, there is an ideal value of $k$ that achieves uniqueness and an estimable model. 
In our experiments, we choose $k$ to optimize performance.

To reduce computational complexity, we further constrain the emission distribution.
As errors are relatively rare, we only allow \kmers within a small Hamming distance of an observed $k$mer $\baseCalls_t$ to have non-zero emission distributions.
If we define the $d$-\emph{neighborhood} of observed $k$mer $\baseCalls_t$ as
    \begin{equation}
        \label{def:kmer_nbhd}
        \nbhd{\baseCalls_t} = \{\boldsymbol{w}: \boldsymbol{w} \in \statespace ~\text{and}~ D(\baseCalls_t, \boldsymbol{w}) \le d \},
    \end{equation}
where $D(\cdot, \cdot)$ is the Hamming distance function,
this assumption reduces the overall number of parameters since $g_t(\indexof{\baseCalls}{t}\mid\Msg_t) \equiv 0$ if $\Msg_t \notin \nbhd{\baseCalls_t}$.


The HMM is fit to the read data using the iterative Expectation-Maximization (EM) algorithm (Baum-Welch).
Subsume all model parameters into vector $\paramSpace$.
The EM locally maximizes the likelihood and produces parameter estimate $\hat{\paramSpace}$.
The initial parameters $\paramSpace^{(0)}$ for the EM are computed as follows.
For transition probabilities, we count the occurrence, $n(\boldsymbol{\alpha}, \beta)$, of $\boldsymbol{\alpha}$ followed by $\beta$ in all reads and set,
    \begin{equation}
        p^{(0)}(\beta|\boldsymbol{\alpha}) = \frac{n(\boldsymbol{\alpha}, \beta)}{\sum_{\beta' \in \Omega} n(\boldsymbol{\alpha}, \beta')}, \boldsymbol{\alpha}\in\statespace, \beta \in \Omega.
    \end{equation}
As for the emission part, we initialize
\begin{equation}
    \begin{array}{rcll}
        q_{tj}^{(0)}(q)		&=& \frac{1}{Q_{max}},	& q \in\{1, \ldots, Q_{\scriptsize\max}\} \\
        g_t^{(0)}(\beta|\beta') &=& \frac{1}{4},	& \beta,\beta' \in \Omega,
    \end{array}\label{eqn:Init}
\end{equation}
where $Q_{\scriptsize\max}$ is the maximum quality score the sequencer can produce, $j\in\{0,1\}$ indicates presence of an error, and $t\in\{k,k+1,\ldots,L\}$ is read position.
We tried multiple random initializations and found the EM insensitive to choice of $\paramSpace^{(0)}$.
\vss
\subsection{Penalized Estimation}
While the HMM reasonably captures the local dependence present in genomic sequences and the error characteristics of modern sequencers, it fails to recapitulate the finite genome length.
To impose our certainty that the vast majority of $k$mers are unique, we use the approximate $l_0$ penalty proposed in~\cite{Alexander2011},
\[
    J(\paramSpace) = \sum_{\boldsymbol{\alpha}\in\statespace, \beta\in\Omega} \frac{\log(1+\transDist{\boldsymbol{\alpha}}{\beta}/\gamma)}{\log(1+1/\gamma)},
\]
that penalizes small transition probabilities and drives them to zero.
Given the set of observed reads $\setRead$, the EM can be adapted to maximize the penalized log-likelihood
\[
    l(\paramSpace|\setRead) - \lambda J(\paramSpace),
\]
where $\lambda$ and $\gamma$ are user specified tuning parameters.
In general, increasing $\lambda$ or decreasing $\gamma$ strengthens the penalty.
Desirable values for $\lambda$ and $\gamma$ could be determined by imposing a level of sparsity consistent with a prior estimate of genome length.
In the experiments reported here, we seek a strong penalty to push small transition probabilities to zero and eventually eliminate \kmers with suspiciously low coverage.
Therefore, we choose $\gamma = 10^{-4}$ and vary $\lambda$ over the rough grid $\{100,150,200,250,300\}$ to optimize performance.

\section{Error Correction Algorithm}
\vs

\noindent Given a fitted HMM for the data, we now discuss the actual error correction algorithm.
One naturally turns to the Viterbi algorithm to estimate the maximum likelihood ``true read" $\Msg$, given the pair $(\baseCalls, \qScores)$.
However, the state space $\statespace$, even for modest $k$ and relatively small genome size $|\cG|$, is formidable and prevents exact Baum-Welch and Viterbi algorithms. Accordingly, we use an approximate Viterbi-like decoding and sequential decoding as discussed below.
\vss
\subsection{An approximate Viterbi Algorithm}
\label{subsec:AppxViterbi}
\noindent For computational reasons, the Viterbi is limited, like the HMM, to only consider true sequences, $\Msg$, constrained by our assumptions on the emission distribution.
While it propagates likelihoods of survivor paths, if a survivor path contains a state $\Msg_t$ differing at more than $d$ locations from $\baseCalls_t$, then state $\Msg_t$ is deemed implausible and the survivor path is not extended. The same restriction is applied during the Baum-Welch parameter estimation described in Section~\ref{subsec:HMM}. We call this decoding method A-Viterbi.
\vss
\subsection{Sequential Decoding on HMM with Fano Algorithm}
\label{subsec:Fano}

\noindent
%
Sequential decoding was proposed as a way to decode convolutional codes (prior to the optimal Viterbi algorithm) \cite{Wozencraft, Fano}. For codes with high constraint lengths, it serves as a good low-complexity alternative to the Viterbi algorithm. In our work we adapt the Fano algorithm for determining the maximum likelihood state sequence in the HMM. Our discussion here is based on the description in \cite{HanChen} (see Fig. \ref{algo:FanoHMM}). In the Fano algorithm at any given stage there is only one active path, where a path is defined to be a sequence of states $\Msg_1, \dots, \Msg_t$ that have a non-zero probability of occurrence. Let the path labels of the predecessor path, the current path, and the successor path be $\bnu_p$, $\bnu_c$, and $\bnu_s$. The corresponding Fano metrics are denoted as $M_p$, $M_c$, and $M_s$.
A given candidate successor path $\bnu_s$, corresponds to appending a new state to $\bnu_c$ such that the new state has a positive probability of being reached from the last state of $\bnu_c$. The probability of choosing $\bnu_s$ as the successor path can be computed from the transition distribution of the HMM; we denote it as $a(\bnu_c, \bnu_s)$ below. Likewise the emission distribution specifies the probability of the emitted base and quality score corresponding to this transition; this is denoted by $\xi_s$ below (to avoid complicated notation). The Fano metrics are updated as follows.
\begin{eqnarray}
\label{eq:Fanometric}
    M_s = M_c \,+\, \log_2 \big[ a(\bnu_c, \bnu_s) \big] \,+\, \log_2 ( \xi_s ) \,+\, B.
\end{eqnarray}
Here $B$ represents the bias whose value is chosen with the purpose that the Fano metric will keep increasing as long as we are on the correct path \cite{HanChen}. 
\begin{figure}
\begin{algorithmic}[1]
\REQUIRE $k$mers obtained from the read sequence, step size $\Delta$, bias $B$, parameters for HMM
\ENSURE corrected read (or path $\bnu^{*}$) and Fano metric $M^{*}$ \\
\textbf{Initialization}: threshold $T = 0$, \, $\bnu_p = dummy$, \, $M_p = - \infty$, $\bnu_c = $ $k$mer in the first stage, $M_c = 0$, stage $t = 1$. \\
\STATE Choose the successor path which has the largest Fano metric based on the transition probability of $\bnu_c$ and emission probability of the $t + k$th base. Denote this path label as $\bnu_s$ and corresponding Fano metric as $M_s$. \label{algo:fwd4best}
\IF {$M_s \geq T$ } \label{algo:msjudge}
\STATE Move one base forward and update \\
       $\bnu_p = \bnu_c , \, M_p = M_c$; $\bnu_c = \bnu_s , \, M_c = M_s$; set $t \gets t + 1$
\IF {$t = L - k + 1$}
\STATE Stop algorithm and output $\bnu^{*} , \, M^{*} = M_c$
\ELSE
\IF {$M_p < T + \Delta$}
\STATE Tighten threshold $T$ such that $T \leq M_c < T + \Delta$.
 Go to step \ref{algo:fwd4best}.
\ENDIF
\STATE Go to step \ref{algo:fwd4best}.
\ENDIF

\ELSE
\IF {$M_p \geq T $} \label{algo:mpjudge}
\STATE Move one base back and update \\
       $\bnu_s = \bnu_c$, $M_s = M_c$; $\bnu_c = \bnu_p$, $M_c = M_p$; $t \gets t - 1$ \\
       and re-compute $M_p$ and $\bnu_p$.
\STATE Attempt to find the non-visited successor path of $\bnu_c$  which has the largest Fano metric. Denote this path as $\bnu_t$ and its Fano metric as $M_t$.
\IF {$\bnu_t$ is empty}
\STATE Go to Step \ref{algo:mpjudge}.
\ELSE
\STATE Update $\bnu_s = \bnu_t, \, M_s = M_t $ and go to Step \ref{algo:msjudge}.
\ENDIF

\ELSE
\STATE Lower threshold as $T \gets T - \Delta$ and go to Step \ref{algo:fwd4best}.
\ENDIF
\ENDIF

\end{algorithmic}
\caption{Sequential Decoding with Fano Metric on HMM}
\label{algo:FanoHMM}
\end{figure}
\section{Experimental Results}
\vs
\label{sec:results}

\noindent
We compared the performance of the A-Viterbi and Fano decoding
algorithms to Reptile on one simulated and one real dataset (Table~\ref{tab:datasets}).
Reptile is a top-performing method in a recent survey of error correction methods~\cite{Yang2012}.
All the results we present here assume we know the first true $k$mer.
Since we use simulation or resequencing experiments of known genomes, we can reliably infer this information.
In practice, a known primer sequence is often attached to both ends of the DNA fragment being sequenced, so the assumption is not restrictive.
All methods include model complexity parameters, such as $k$ and $d$, that can be difficult to choose when the true sequence is unknown.
In our experiments, with the true sequence available, we can tune these parameters for optimal performance.
We emphasize that we have tuned \textit{all} methods to achieve their respective best performance.

For each dataset, maximum likelihood estimates of the HMM parameters were estimated using the penalized likelihood ($\gamma=0.0001, \lambda=250$) for various \kmer\ lengths, $k=13, 14, 15$, and maximum Hamming distance, $d=4$.
Then, A-Viterbi and Fano were used to perform the decoding at each chosen $k$.
The A-Viterbi algorithm was run using the same $d$ used to estimate the HMM parameters.
For the Fano algorithm, parameter $B$ in Eq.~(\ref{eq:Fanometric}) was set to be $2$ and $10$ for the simulated and real dataset, respectively; we tried both $\Delta = 0.5$ or 1.0.
Note, the Fano algorithm places no constraints on the \kmer Hamming distances.

Reptile uses a ``tile'' formed by concatenating two \kmers\ of length $k$ with overlap of length $k - \mbox{\it step}$.
Corrections are made if the observed tile is uncommon and there is a substantially more common tile in the neighborhood of the observed tile.
The tile neighborhood is formed by allowing up to $d$ errors in each $k$mer.
Tile counts are computed from high quality reads only.
We chose the best parameters $(k, \mbox{\it step})$ using a grid-search over $7 \le \mbox{\it step} \le k \le 12$.
Given $(k,\mbox{\it step})$, thresholds for error correction decisions were automatically selected, following the instructions in the software manual and \textit{MaxBadQPerKmer} was left at the default $4$.
The maximum errors allowed per \kmer was set to $d=4$.
All remaining parameters were left at their defaults.

Let $e$ be the total number of ground truth errors in the sequencing reads excluding those in the first \kmer (or tile for Reptile).
The probability of error correction is defined as $\zeta \triangleq ce / e$ and gain is defined as $\eta \triangleq (ce -fa) / e$, measuring the effective number of errors removed from the dataset~\cite{Yang2010}.

\begin{table}[!t]
\centering
\caption{Benchmark sequencing datasets}
\label{tab:datasets}
\begin{tabular}{c|c|c|c|c|c}
    \hline
    \multirow{3}{*}{Dataset} & Genome	 & Read    & Number  &           & Error \\
                             & length    & length (bp)  & of reads & Coverage & rate (\%) \\
    \hline 
    D1 & 250000 & 36 & 1000000 & 144.0x & 1.23 \\
    D2 & 500000 & 36 & 2132517 & 153.5x & 0.51 \\
    \hline
\end{tabular}
\vss\vss\vss\vss
\end{table}
\vss\vss
\subsection {D1: Simulated Dataset}
\label{subsec:TestGenome}
To create the simulated dataset, one million reads were generated by randomly sampling 36bp sequences from a 250Kbp region (1000Kbp --- 1250Kbp) of the \emph{E. coli} genome (Accession NC.000913).
From the real dataset (see~\ref{subsec:TestReal}), we estimated empirical distributions of quality scores given read positions,
and used these to generate quality scores for every position of each simulated read.
We assumed the simulated quality scores indicated the true error probabilities and replaced the true base $\beta_t$ at position $t$ with error base $\beta_t'\neq \beta_t$ with probability $\frac{10^{-q_t/10}}{3}$ where $q_t$ is the quality score.

Table~\ref{tab:d1} shows the error correction results for various choices of \kmer length, $k$, or $(k,\mbox{\it step})$ for Reptile.
In bold, we show the best performance for each method, as measured by the gain metric $\eta$ along with the results for a few additional, nearby settings.
Both Fano and A-Viterbi outperform Reptile by achieving higher error correction probabilities while having a lower false alarm probability.
Overall, the Fano algorithm is the best performer at $k = 15$ and $\Delta = 0.5$.

The HMM-based approaches and Reptile exhibit best performance at different values of $k$,
but Reptile is much more sensitive to this choice.
In a reference-free error correction setup, when no explicit ground truth is available to guide choice of $k$, the HMM-based approaches have the advantage of yielding robust performance over a wider range of $k$.
Quake~\cite{Kelley2010} recommends choosing $k$ such that $\frac{2|\cG|}{4^k}\approx0.01$, which suggests $k=13$ in this case.
While Fano and A-Viterbi are not at their peak performance for this choice of $k$, their performance is near-optimal.
In contrast, the Reptile authors recommend $k = \log_4 |\cG| \approx 9$, which indeed works well.



\begin{table}[!t]
\renewcommand{\arraystretch}{1.3}
\centering
\caption{Error Correction Results for D1}
\label{tab:d1}
\begin{tabular}{c|c|c|c|c|c|c}
\hline
& & $k^\dagger$ & $ce$ & $\zeta$ & $fa$ & $\eta$ \\
\hline
\multirow{3}{*}{Fano}
&  & 13  & 430391 & 0.9981 & 100 & 0.9979 \\
& $\Delta$ = 0.5 & 14  & 428324 & 0.9993 & 21 & 0.9993 \\
&  & {\bf 15}  & {\bf 425442} & {\bf 0.9996} & {\bf 6} & {\bf 0.9996} \\
\hline
\multirow{3}{*}{A-Viterbi}
&  & 13 & 430384 & 0.9967 & 227 & 0.9962 \\
&  & 14 & 427839 & 0.9978 & 102 & 0.9975 \\
&  & {\bf 15} & {\bf 424881} & {\bf 0.998} & {\bf 75} & {\bf 0.9979} \\
\hline
\multirow{2}{*}{Reptile}
& & (8, 8) & 355395 & 0.8423 & 3900 & 0.8331 \\
& & $\text{\bf (9, 9)}$ & {\bf 405971} & {\bf 0.9848} & {\bf 678} & {\bf 0.9832} \\
& & (10, 10) & 314306 & 0.7867 & 708 & 0.7849 \\
\hline
\end{tabular}

\vspace{3pt}
{\footnotesize $\dagger:~$ For Reptile, this column is reported as $(k, step)$.}
\vss\vss\vss\vss
\end{table}

\vss
\subsection{D2: Real Experimental Dataset}
\label{subsec:TestReal}
To test the performance of our model on a real Illumina dataset, we used the data of an \textit{E. coli} resequencing experiment (Accession SRX000429).
Knowing the reference genome allows us to identify the ``ground truth'' errors as long as we can identify the position of the read in the reference genome. 
To align the reads to the reference genome, we used the Burrows-Wheeler Aligner~\cite{Li2010} with default parameters. 
We selected all reads with a unique match to a 500Kbp region (1000Kb --- 1500Kbp) on the reference genome and tallied the true errors as mismatches between the selected reads and the reference sequence.
There are cases of erasure where the nucleotide is recorded as ``N'' for ``not determined'' in the reads.
For Reptile and A-Viterbi, all $N$ bases were replaced by $A$, but were left intact for Fano, which can directly correct base $N$.

\begin{table}[!t]
\renewcommand{\arraystretch}{1.3}
\centering
\caption{Error Correction Results for D2}
\label{table:d2}
\begin{tabular}{c|c|c|c|c|c|c}
\hline
& & $k$ & $ce$ & $\zeta$ & $fa$ & $\eta$ \\
\hline
\multirow{3}{*}{Fano}
& & 13  & 354864 & 0.9341 & 7866 & 0.9134 \\
& $\Delta$ = 0.5 & \bf{14} & \bf{353209} & \bf{0.9397} & \bf{5589} &  \bf{0.9248} \\
& & 15 & 348592 & 0.9388 & 5348 & 0.9244 \\
\hline
\multirow{3}{*}{A-Viterbi}
 &  & 13 & 357945 & 0.921 & 6950 & 0.9026 \\
 &  & {\bf 14} & {\bf 350316} & {\bf 0.9172} & {\bf 5076} & {\bf 0.9036} \\
 &  & 15 & 344943 & 0.9152 & 4756 & 0.9024 \\
\hline
\multirow{3}{*}{Reptile}
 & & (8, 8) & 270226 & 0.7461 & 39280 & 0.6377 \\
 & & {\bf (9, 9)} & {\bf 314154} & {\bf 0.8935} & {\bf 20748} & {\bf 0.8345} \\
 & & (10, 10) & 250480 & 0.7973 & 8924 & 0.7130 \\
\hline
\end{tabular}
\vss\vss\vss\vss
\end{table}

The error correction results on dataset D2 are summarized in Table \ref{table:d2}.
The superiority of the Fano algorithm is accentuated in this real data.
We presume that Fano outperforms A-Viterbi by exploring parts of the 4-ary tree ruled out by A-Viterbi due to the neighborhood constraint (cf. Section \ref{subsec:AppxViterbi}).
The Quake-recommended $k$ for this dataset is $14$; Reptile recommends $9$.
\vss\vss\vss
\subsection{Parameter Selection and Sensitivity}
While not as sensitive as Reptile to choice of $k$, the HMM-based methods are sensitive to other parameter choices.
We mention only the strongest effects here as we have not yet completed a thorough analysis.
Most striking, A-Viterbi achieved gain of only $0.857$ with $\gamma=0.001$ and $\lambda=500$.
The Fano algorithm is most sensitive to $B$. In the real dataset, the HMM parameters were such that a larger value of the bias $B$ needed to be chosen; otherwise there was excessive backtracking in the running of the algorithm. 

Since we had roughly optimized the A-Viterbi and Fano algorithms over $\lambda,\gamma,\Delta$, and $B$, we attempted to optimize Reptile over two of its parameters.
Through much experimentation, we could improve Reptile performance to $\eta=0.9234$ when \textit{T\_expGoodCnt}$=28$ and \textit{T\_card}$=13$ on the real dataset, which is very close yet marginally inferior, to the best Fano performance.
However, neither HMM-based method was allowed quite the same diligent exploration of its parameter space.
It is clearly imperative that we develop methods to choose $\lambda, \gamma, \Delta$, and $B$ without reference to a test genome, so all methods can be compared on equal footing. Our current set of experiments leads us to conclude that HMM-based methods are substantially easier to tune, less sensitive to parameter settings, and more flexible when it comes to adopting more realistic emission distributions.

\vss
\section{Conclusion}
\vs
\label{sec:conclusion}

We establish the HMM for the problem of noisy DNA read correction and develop the approximate Viterbi algorithm and the sequential decoding algorithm with Fano metric to execute the error correction. Based on our test results for both simulated and real data, the proposed algorithms often outperform another state of the art method in this field. Our future work will include development of systematic procedures for choosing the parameters of the algorithms, an investigation of more complex, context dependent error emission distributions and an exhaustive comparison with respect to competing methods.


\section*{Acknowledgment}
\vs
This work was funded in part by NSF awards DMS-1120597 and CCF-1149860.


\bibliographystyle{IEEEtran}   
\bibliography{IEEEabrv,ISIT2013}   

\end{document}